# "Smarter" NICs for faster molecular dynamics: a case study

Sara Karamati*, Clayton Hughes†, K. Scott Hemmert†, Ryan E. Grant‡, W. Whit Schonbein†,
Scott Levy†, Thomas M. Conte*, Jeffrey Young*, Richard W. Vuduc*
*Georgia Institute of Technology, Atlanta, Georgia, USA
†Sandia National Laboratories, Albuquerque, New Mexico, USA
‡Queen's University, Kingston, Ontario, Canada
*Email: {s.karamati,jyoung9,richie}@gatech.edu

*Abstract*—This work evaluates the benefits of using a "smart" network interface card (SmartNIC) as a compute accelerator for the example of the MiniMD molecular dynamics proxy application. The accelerator is NVIDIA's BlueField-2 card, which includes an 8-core Arm processor along with a small amount of DRAM and storage. We test the networking and data movement performance of these cards compared to a standard Intel server host using microbenchmarks and MiniMD. In MiniMD, we identify two distinct classes of computation, namely core computation and maintenance computation, which are executed in sequence. We restructure the algorithm and code to weaken this dependence and increase task parallelism, thereby making it possible to increase utilization of the BlueField-2 concurrently with the host. We evaluate our implementation on a cluster consisting of 16 dual-socket Intel Broadwell host nodes with one BlueField-2 per host-node. Our results show that while the overall compute performance of BlueField-2 is limited, using them with a modified MiniMD algorithm allows for up to 20% speedup over the host CPU baseline with no loss in simulation accuracy.

## I. INTRODUCTION

SmartNICs are network interface cards with extended computational capabilities (see Section II-A and the survey by Grant et al. [1]). There is growing agreement that the ability to do on-NIC compute will have a critical enabling role in cloud and datacenter architectures, especially in tackling functions such as networking control, storage management, and security. On-NIC compute capabilities might take the form of special-function ASICs or FPGAs, but in this paper, we are especially interested in the case of embedded *general-purpose* multicore processors and memories. We focus on the NVIDIA (née Mellanox) BlueField SmartNIC designs, with experiments conducted on the BlueField-2 implementation.

Having a flexible compute unit close to both the host CPU as well as the network infrastructure raises a number of questions about the role they can play in other classes of applications, such as those in high-performance computing (HPC). While exploratory work along these lines exists, especially at the middleware layer (Section II-A), applications-level work remains nascent. Thus, in this paper we investigate the potential of SmartNICs as a de facto compute accelerator for HPC applications. That is, are there other possible scenarios to use SmartNICs besides previously assumed tasks like networking control, storage management, or security?

We consider this question using the case study of MiniMD, a molecular dynamics (MD) simulation benchmark. MiniMD is simple enough to study in detail while also having challenging characteristics: the simulation behavior and its communication to computation ratio can vary with different input and configuration-parameter values, including problem size, the number of processors, and the re-neighboring frequency, among others, making it an excellent choice to examine the new hardware in some detail.

There are three key challenges to porting MiniMD to BlueField profitably. First, there is a sequential dependence between the core parts of each iteration of MiniMD, making it difficult to concurrently offload communication- or computation-related routines to BlueField. Second, for larger problem sizes, MiniMD's communication time is small compared to its computation time, making it nontrivial to extract reasonable performance by hiding communication load. Third, our experiments with the OSU MPI microbenchmark in Section III show that BlueField does not actually outperform conventional host-to-host communication in latency or bandwidth due to its slow CPU.

To overcome these challenges, our strategy is to change the algorithm. The original MiniMD algorithm relies on an expert-tuned parameter-based heuristic to periodically rebuild and maintain certain internal data structures, which suggests alternative heuristics are possible. We develop one that can relax the sequential dependence of certain operations in MiniMD. This change exposes additional task-parallelism that a BlueField can then exploit. This approach is an instance of how platform characteristics can inform a redesign of the application's algorithm and implementation.

We evaluate our method experimentally against the MiniMD baseline on the Thor cluster at the HPC Advisory Council (Table I). Our contributions are:

- We conduct performance analysis to understand the opportunities and limitations presented by the BlueField for potential HPC applications.
- We evaluate strategies for partitioning communication- and compute-oriented tasks for the MiniMD application. The potential offloading scenarios considered are 1) communication-heavy routines offloaded to the Blue-Field, and 2) computation-heavy routines offloaded to

BlueField. The former is "natural" while the latter would be "unexpected" due to the relatively weak performance of BlueField cores. Nevertheless, we show, surprisingly, that the latter can outperform the former.
- Motivated to exploit the BlueField, we restructure the algorithm to decouple an infrequent periodic-update heuristic from the algorithm's main loop, exposing more task parallelism that the BlueField can be used to help execute.
- We evaluate the efficiency of our proposed offloading method and show that the performance can be boosted up to 20% above the original baseline. Moreover, it appears that this increase in performance is possible with only a modest increase in node-level power cost (6% to 13% increase in power as a rough estimate). We also construct performance models to estimate the theoretical maximum performance improvement possible by offloading all communication routines to BlueField to explain our results.

Taken together, these findings constitute a positive result on the potential of SmartNIC co-processing and pave the way for future studies that consider other computational motifs in HPC, as we suggest in Section VI.

## II. BACKGROUND

Two pieces of background are helpful for understanding our work: 1) a discussion of the BlueField and related work, which helps to contextualize our study (Section II-A), and 2) a basic description of how our benchmark application, MiniMD, works (Section II-B).

### A. BlueField Data Processing Units

The class of SmartNIC platforms we are targeting implement their on-board "smarts" using general-purpose multicore processors and NIC-private memories. This class sometimes goes by the moniker of *data processing units*, or DPUs. The specific DPU of interest in our project is the NVIDIA (née Mellanox) BlueField, which uses an Arm multicore processor.

DPUs are distinct from other SmartNIC platforms, which might instead rely on packet-processing ASICs or FPGAs [2]–[7]; for an explanation of these design choices, see the excellent survey by Grant et al., which broadly catalogs the challenges and opportunities for SmartNICs [1]. In our project, we are asking what implications DPUs have for the implementation of high-performance scientific computing algorithms.

By way of contrast, another natural way to use SmartNICs is to offload specific middleware operations. A compelling example for BlueField is BluesMPI, which implements MPI's all-to-all exchange for BlueField [8]. The designers of BluesMPI identify the most promising regime for acceleration as being medium to large messages in a nonblocking all-to-all. They show that their drop-in all-to-all replacement can reduce the bottleneck of large three-dimensional fast Fourier transforms. This demonstration shows a path of least resistance to accelerating real HPC applications with minimal or no changes, since any changes are isolated within the middleware layer.

For more general applications, there are two basic approaches to using the computational cores of a DPU: an *in-pipeline* or "on-path" model, in which we inject custom computation on message data between their receipt from the network and delivery to the host, versus an *asynchronous* or "off-path" model, in which we run arbitrary computation concurrently with both communication and any on-host computation. Earlier work in INCA [9] provides an example of a hybrid approach by implementing asynchronous off-path style compute operations onto on-path networking hardware.

Liu et al. observe that the latency of RDMA primitives exposed in off-path SmartNICs can take twice as long as that of native blocking DMA in on-path SmartNICs [10]. Furthermore, by fully utilizing PCIe bandwidth, non-blocking DMA primitives of on-path SmartNICs perform even better than blocking ones, regardless of message size. Liu et al. also show that the hardware traffic manager in on-path SmartNICs provides an additional performance advantage by reducing the synchronization cost through a shared queue abstraction when multiple cores need to pull incoming packets from this queue. Based on these experimental observations, they propose an actor-based framework for offloading distributed applications onto SmartNICs, showing that it decreases host CPU utilization and lowers application latency for on-path SmartNICs.

Another recent study of on-path operations on BlueField provides a detailed characterization of the potential to overlap computation and communication [11]. One finding is that the BlueField's processor is weak enough that, unlike typical x86 host cores, using all BlueField cores cannot fully saturate available link bandwidth in a 100 Gbps network, maxing out instead at around 60%. And even at that level, only about 70% or so of the processor's compute capability is available for in-pipeline operations. But a different finding is that the BlueField cores can be competitive or better than x86 processors of just one or two generations earlier. In particular, vectorizable math operations perform well, opening an avenue to consider scientific computing applications.

One such application is the PENNANT proxy app, which Williams et al. study on both first and second generation BlueField DPUs [12]. Their results are pessimistic, as they were unable to achieve any appreciable speedups. However, they also deliberately limit their code transformations to the kind of incremental offloading one might expect in a first-attempt to exploit DPU processing capabilities. They leave open the possibility of more aggressive algorithmic and software restructuring, which might yield better results. Our study pursues such restructuring transformations for MiniMD.

Jain et al. explore BlueField offloading for different phases of deep learning training, i.e., data augmentation, model validation, or both [13]. Their proposed offloading schemes can speed up training by up to 15%. However, they did not find one offloading scheme that could optimally speed up training all the different models in their study. Specifically, the performance gain is sensitive to the amount of overlap in computation that a given offloading scheme provides between the host and SmartNICs for the model under training. This observation illustrates the difficulty of attaining performance improvements from off-path techniques.

One significant challenge to aggressive restructuring is programmability. There exist complementary lines of research and development that seek to make open, portable interfaces for SmartNICs and DPUs. An example is OpenSNAPI [14]. While its use is outside the current scope of our study, we view OpenSNAPI as a critical enabling technology that could simplify future work like ours.

*B. MiniMD*

MiniMD is a benchmark from the Mantevo benchmark suite that implements a molecular dynamics (MD) simulation [15]. It is a proxy application for LAMMPS [16]. Both LAMMPS and MiniMD provide simulation capabilities for particles in solid, liquid, or gaseous media, but MiniMD only supports modeling the effects of a Lennard-Jones (LJ) potential between pairs of molecules and many-body interactions via the Embedded Atom Model (EAM).

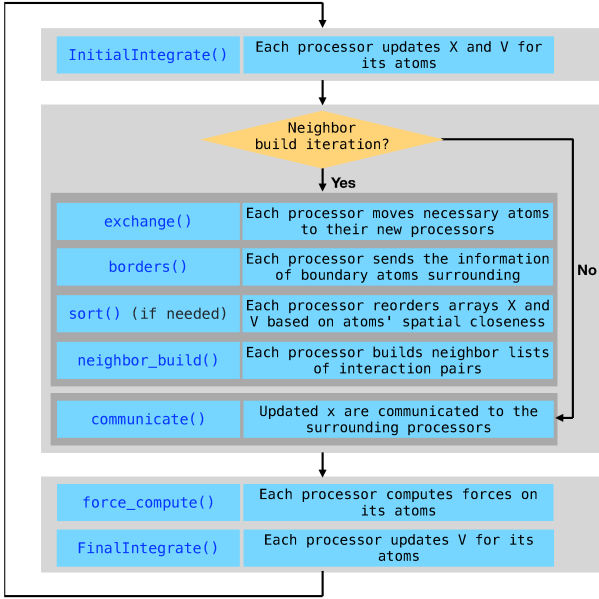

Fig. 1. MiniMD flow diagram.

The flow of operations in one iteration of MiniMD appears in Fig. 1. The core algorithm consists of computing the force exerted on each atom from those within a fixed neighborhood (i.e., **force_compute**()), followed by updating the position and velocity of atoms due to that force (i.e., **InitialIntegrate**() and **FinalIntegrate**()). Furthermore, for each atom, the algorithm maintains a neighbor list to help quickly determine which pairs of atoms interact (**neighbor_build**()). This list consists of all the atoms within a cut-off distance $r_c$, plus additional atoms within an extra buffer distance (also known as the skin distance) $\Delta$. To save computation time, the algorithm only updates the neighbor lists every $n$-th iteration, where $n$ is the *re-neighboring frequency*, an input to the algorithm (Fig. 2). These parameters are illustrated in Fig. 2.

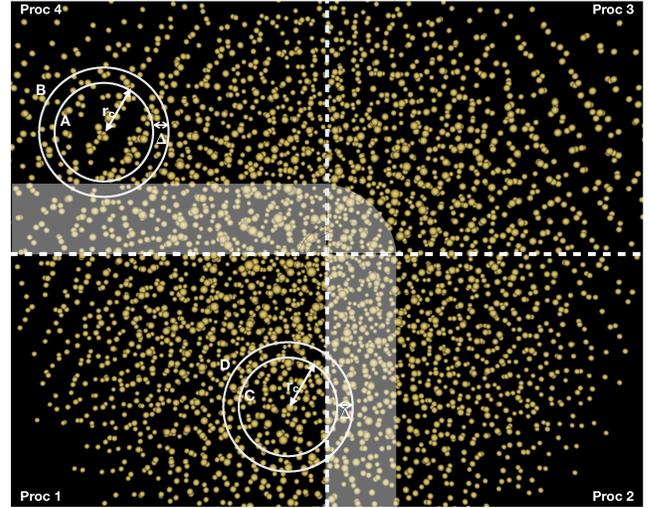

Fig. 2. Illustration of a sample MiniMD simulation in two dimensional space; yellow dots are particles, the dotted lines indicate the boundaries of the spatial domain of each processor (assuming 4 processors working on the problem); circles A and C contain particles within cut-off distance $r_c$; circles B and D contain particles within an extra buffer distance $\Delta$. Circle D contains particles owned by a neighboring processor and these particles are in the halo region of the proc 1 (gray box); Particle information within the halo region of proc 1 must be communicated to proc 1 at every iteration.

Every iteration, each atom interacts only with neighbors that are less than $r_c$ away. To be able to reduce the neighbor list update frequency, the algorithm uses the atoms tracked in the skin of each atom as the atoms move after each time step. The MiniMD proxy uses the same list in subsequent iterations until the next iteration in which the neighbor lists must be updated. Additionally, the atoms within a processor's subdomain may be reordered in memory at regular intervals in order to improve cache performance. (This reordering is implemented using a sorting operation.)

MiniMD exploits parallelism by dividing the spatial domain of the simulation into partitions and assigning each partition to a process. In addition, each process designates part of its partition as its boundary region, which is also decided by the skin distance. The main iteration consists of the following communication routines to coordinate between processes:

- **exchange**(), where each process checks the location of its atoms, sends information about the atoms that have moved outside its partition to a neighboring partition on another processes, and receives the information for atoms that have entered its partition from other processes. This routine is invoked only when neighbor lists are updated.
- **border**(), where each process creates a list of atoms in its boundary region that may fall in the neighbor lists of the atoms in the surrounding partitions. Additionally, **border**() sends the list of these atoms and their positions to the corresponding adjacent processes. This routine is invoked only when neighbor lists are updated.

- **communicate()**, where the updated positions of atoms in the boundary region are sent to the corresponding processes. This information is similar to what **border()** sends out to the adjacent processes. This routine is invoked during the iterations where neighbor list is not being updated.

A key observation about Fig. 1 is that when a neighbor-building iteration occurs (middle of figure), it must precede the next round of force computations (bottom). This serialization would, at first glance, appear to preclude computation-communication overlap.

## III. MOTIVATING EXPERIMENTS

We first evaluate the BlueField-2 using the OSU Microbenchmarks suite (OSU version: 5.7.0) [17] and the "off-the-shelf" version of MiniMD. This evaluation seeks to understand the opportunities and limitations presented by the BlueField DPU for the unmodified baseline. These findings then inform our approach of Section IV.

### A. Experimental setup

Our experimental platform is the Thor cluster within the HPC-AI Advisory Council Testbed, which contains MBF2H516A-EENOT Full-Height Half-Length (FHHL) DPUs.[1] Thor is a 32-node cluster containing dual-socket Intel Xeon 16-core Broadwell-class CPUs running at 2.60 GHz (Host) and a BlueField data processing unit. Each DPU combines the ConnectX-6 Dx HDR100 100 Gbps InfiniBand/VPI adapters with 8 ARMv8 A72 cores operating at 2.5 GHz. Additional configuration details appear in Table I.

We use NVIDIA's HPC-X Rev 2.8.1 Software Toolkit, including OpenMPI version 4.1.2a1 to build and run the codes on hosts and BlueField devices [18]. For many of the motivating experiments of this section, unless otherwise specified, we assign one MPI process to an entire node. This configuration is in contrast to assigning each core to a different MPI process or enabling multithreading within a node, which we do consider in Sections III-D and V.

All performance metrics are collected using MiniMD's default configuration, which employs the Lennard-Jones potential. In our experiments, we vary the number of atoms and re-neighboring frequency. Each experiment is run for 1,000 iterations.

### B. OSU Microbenchmarks

We ran microbenchmarks designed to assess the performance of the elements of MPI-based applications that occur in MiniMD:

- Point-to-point latency (osu_latency).
- Point-to-point bandwidth (osu_bw).
- Point-to-point multi-pair bandwidth and message rate (osu_mbw_mr).
- Collective allgather latency (osu_allgather).

[1]See: https://www.hpcadvisorycouncil.com/cluster_center.php

For the first three tests, we compare the network performance for three different arrangements: internode communication between hosts (Host-to-Host), internode communication between BlueFields (BF-to-BF), and internode communication between a host and its corresponding BlueField (Host-to-BF).

For collective and multi-pair tests, we compare the Host-to-Host and BF-to-BF arrangements. For all tests, we report the normalized results with respect to Host-to-Host performance. Figure 3 shows the results from the OSU microbenchmark latency and bandwidth tests. They indicate that the latency between two BlueFields is higher for smaller message sizes than the latency between two hosts, but there is a clear transition point (16 KiB) beyond which the BF-to-BF performance improves. This transition is due to the shift from eager to rendezvous protocol at message sizes above 16 KiB. Unlike the rendezvous protocol, the eager protocol directly copies the data, which significantly affects BF point-to-point performance due to its slow CPUs. For the same reason, we observe a similar trend for multi-pair bandwidth and message rate tests (Section III-B) and collective allgather latency (Fig. 5). The multi-pair bandwidth and message rate test is performed between 8 pairs of processes with a window size of 64. The collective allgather test is run over 16 nodes.

Contrary to our expectations, we observe degraded performance with the BlueField despite being "near-network," especially for smaller message sizes. These observations suggest that only offloading communication load to BlueField is unlikely to improve performance, suggesting alternative strategies like off-path co-processing.

### C. Experimental Analysis of MiniMD

To gain more insight into the performance of the BlueField, we compare the performance of parallel MiniMD on host CPUs and BlueField Arm CPUs. Figure 6 shows the slowdown of MiniMD on BlueField compared to the corresponding host-only run. It shows that the observed slowdown is between 1.64 and 2.53. The magnitude of slowdown depends on the ratio of computation to communication. In particular, we can see that the slowdown is more pronounced for larger problem sizes and smaller numbers of processors. In MiniMD, as the size of the problem increases or the number of processors decreases, the size of the spatial box allocated to each processor becomes much greater than the cutoff distance. Consequently, each processor spends more time running computations on its atoms while communicating a relatively small amount of data with adjacent processors [19]. These trends indicate that the increased share of computation causes a measurable performance degradation.

Figure 6 also shows that increasing the re-neighboring interval causes further slowdown on BlueField compared to the host-only run. This performance degradation is because of the added share of computation load relative to communication in each re-neighboring iteration. This relative difference in communication and computation becomes even more prominent for larger problem sizes.

TABLE I
EXPERIMENTAL SYSTEM CONFIGURATION. THE TESTBED IS A 32-NODE CLUSTER, WHERE EACH NODE CONTAINS A DUAL-SOCKET x86 HOST AND ONE
BLUEFIELD. EACH ROW OF THE TABLE BELOW IS A PER-NODE CONFIGURATION. THE LINK BANDWIDTH IS 12.5 GB/s (INFINIBAND HDR AT 100 Gbps).

| Host | Sockets x CPU | Cores per socket | Peak flop/s per socket | Memory | Peak GB/s per socket | Device Type |
|---|---|---|---|---|---|---|
| Thor | 2 × Intel Broadwell (E5-2697A), 2.6 GHz | 16 | 656.6 Gflop/s | 256 GiB | 76.8 GB/s | Host CPU |
| ThorBF | 1 × Arm A72, 2.5 GHz | 8 | 80.0 Gflop/s | 16 GiB | 25.6 GB/s | BlueField P-Series |

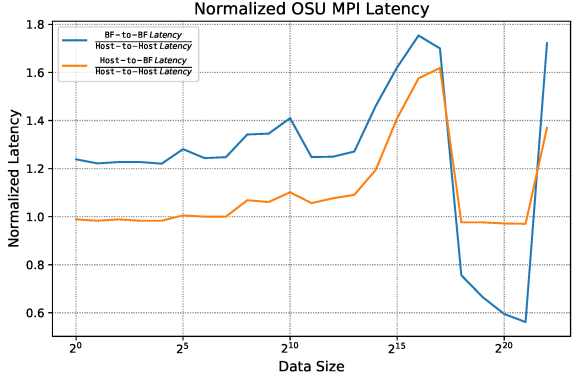
(a) OSU MPI latency

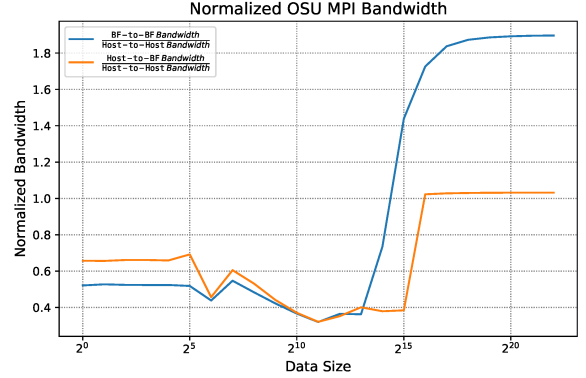
(b) OSU MPI bandwidth

Fig. 3. OSU MPI latency and bandwidth tests, relative to conventional host-to-host communication: BF-to-BF latency is higher, and bandwidth lower, than host-to-host communication for message sizes under 16 KiB. (Data sizes are in bytes.)

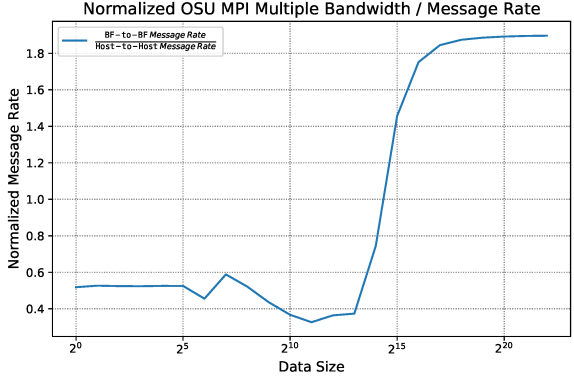
(a) OSU MPI multiple message rate

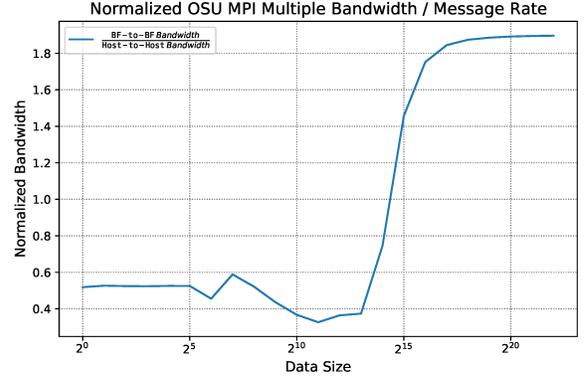
(b) OSU MPI multiple bandwidth

Fig. 4. OSU MPI multiple message rate and multiple bandwidth tests, run between 8 pairs of BF-to-BF or host-to-host processes: Like Fig. 3, a crossover around 16 KiB occurs when BF-to-BF communication outperforms host-to-host communication. (Data sizes are in bytes.)

The execution time breakdown of MiniMD, in the host-only setting, appears in Fig. 7. Here, $t_{\text{total}}$ is the overall execution time, $t_{\text{force}}$ is the time consumed by the **force_compute()** routine, $t_{\text{neigh}}$ is the time consumed for the **neighbor_build()** routine, and $t_{\text{comm}}$ is cumulative time spent on the **exchange()**, **border()**, and **communicate()** routines. The time $t_{\text{comm}}$ is not pure communication time; it also includes the time required to prepare the data for communication. We can see that $t_{\text{comm}}$ has a small share of the overall execution time. Therefore, in a host-BlueField hybrid setting, offloading only the communication routines to BlueField would not result in a significant overall performance gain. Additionally, since MiniMD's communication tasks depend on prior computation steps, decoupling these routines from the rest of the application and offloading them to a co-processor, while achieving full computation-communication overlap, is not a trivial task.

So what could be done instead? Figure 7 indicates that additional computation overlap may be possible. This finding motivates our design approach in Section IV, which seeks to offload work to BlueField.

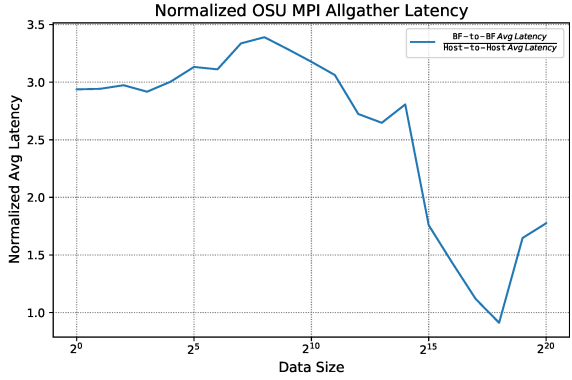

Fig. 5. OSU MPI Allgather latency: The latency of a BF-to-BF allgather operation among 16 BlueFields is never much faster than the equivalent host-to-host allgather among 16 hosts. (Data sizes are in bytes.)

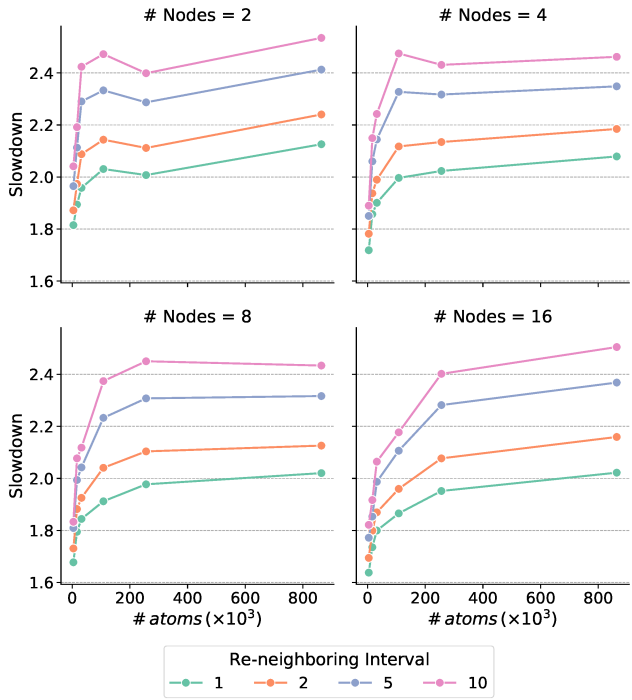

Fig. 6. MiniMD performance slowdown on BlueField compared to the corresponding host-only run: Due to comparatively weak processors, a BlueField-only run can take over twice as long as the corresponding host-only run.

### D. BlueField as a "standalone" platform

Although a single BlueField appears underpowered compared to its host, it also has a lower flop:byte ratio than the host, which arguably makes it an efficient, well-balanced building block. In a "match-up" of hosts against BlueFields, it would take about 16 BlueFields to match the peak of one host using the metrics in Table I:

$$\frac{(\text{host peak flop/s per socket}) \times (\text{\# sockets})}{(\text{BlueField peak flop/s})} \approx 16.4 \quad (1)$$

We can check this effect experimentally. In particular, we first run MiniMD with a problem size of 864,000 particles on a *single* host node but with full concurrency, i.e., 32 OpenMP threads to utilize all of its cores. Then, we run MiniMD using *only* BlueFields, with one MPI process per BlueField and varying numbers of OpenMP threads (1 to 8). The results appear in Fig. 8. When matched on theoretical peak (16 BlueField with full 8-way intra-BlueField concurrency, against a fully concurrent single-host run), the BlueField-only run is over $1.75\times$ *faster*.

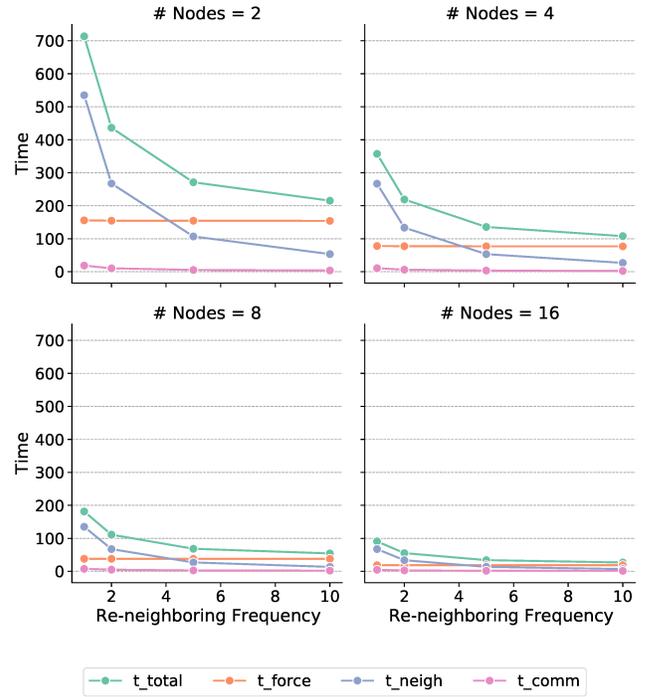

Fig. 7. MiniMD running time breakdown: The time spent in communication is modest under most circumstances. Thus, accelerating only communication operations is unlikely to improve performance, thus motivating our strategy to seek other forms of computational offload.

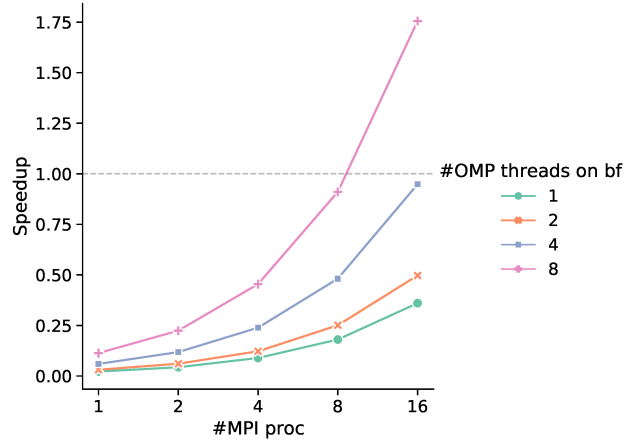

Fig. 8. MiniMD performance for different configurations of BlueField nodes compared to the performance of one host node with 32 cores

## IV. ALGORITHM DESIGN

Armed with the preliminary results of Section III, we explore different ways of reorganizing MiniMD to exploit the presence of a BlueField SmartNIC.

The most natural opportunity is to overlap a call to **neighbor_build** with **force_compute** on the previous neighbor list. For example, consider the execution timeline of baseline MiniMD as illustrated in Fig. 9 (a). Each neighbor-list build step ("NB") produces new neighbor lists, which are then consumed by the three force-computation steps ("FC") that follow. That is, the baseline serializes the NB and FC steps. Suppose instead that we induce an overlap of NB and FC as shown in Fig. 9 (b). This scheme is different from the baseline. The NB step and the first FC step run concurrently and *both* use the *old* neighbor list. Only the subsequent two FC steps use the new neighbor-list that the NB step produced. The interval between two neighbor-list updates is similar to the original algorithm (i.e., every $n$ iterations), but the onset of changes in neighbor lists used for each computation may shift by one iteration.

This transformation will *not* reproduce bitwise identical results to the baseline, so a question will be whether a tolerable accuracy is still preserved under this new heuristic. However, considering that the original algorithm is itself based on an expert-tuned parameter-based heuristic, we might regard our scheme as simply an alternative heuristic—albeit one that tries to expose more task-level concurrency that the underlying heterogeneous hardware can exploit.

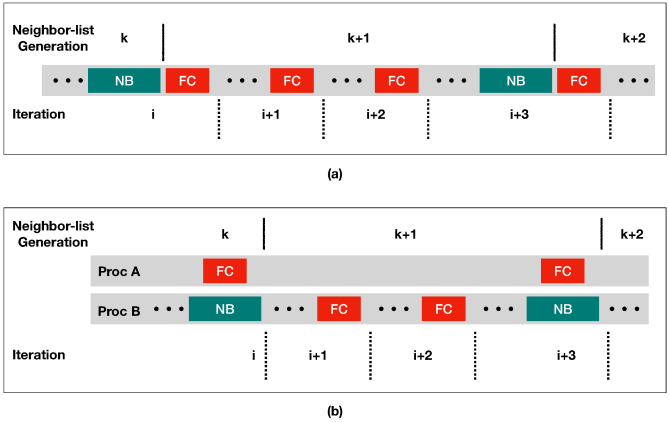

Fig. 9. A schematic of the execution timeline in MiniMD: (*a*, top) The sequence of neighor-list reconstruction ("NB") and force computation ("FC") steps in the baseline implementation of MiniMD. Each NB step precedes one or more FC steps that depend on it. (*b*, bottom) Our goal is to induce overlap of NB and FC steps, through algorithmic and implementation restructuring. One can then imagine offloading either the NB or the FC steps to an available SmartNIC.

There is another catch: it is nontrivial to merge the outputs of the force-computation and neighbor-build steps that run in parallel, due to the design of MiniMD's data structures. To understand why, consider these data structures and what they represent. In MiniMD, each processor keeps its atoms' positions and velocities in arrays $X$ and $V$ where $X[i]$ and $V[i]$ are the position and velocity of the $i$-th atom. After force computation, the accumulated force on the $i$-th atom is stored in the $i$-th location of array $F$ (i.e., $F[i]$). The fact that this specific atom is recognized as the $i$-th atom is purely coincidental, and there is no inverted index maintained to keep track of the future memory index associated with this specific atom in $X$ and $V$ arrays. Therefore, we will refer to $i$ as the temporary identity of this particular atom, until the next neighbor-build routine shuffles the atoms around and likely changes this identity. Keeping the fragility of this temporary identity in mind, MiniMD uses these indices to construct and maintain its neighbor lists.

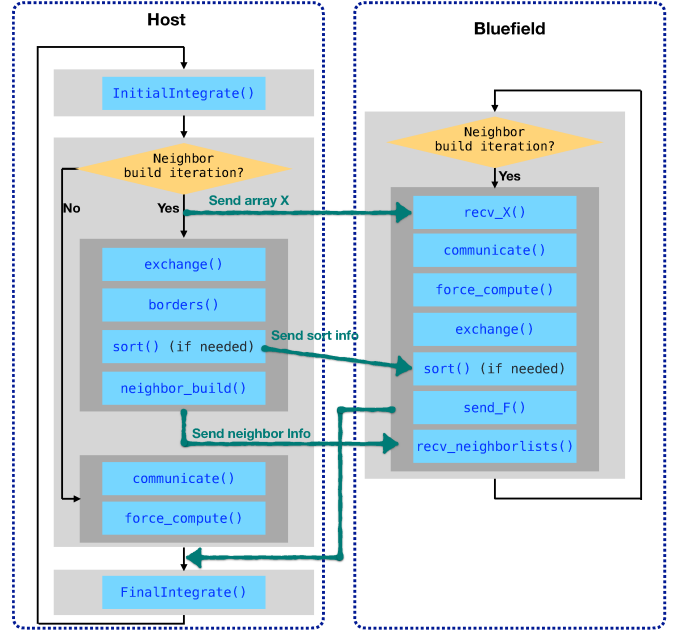

Fig. 10. MiniMD BlueField overlap diagram. During neighbor-build iterations, the host rebuilds the neighbor lists while the BlueField performs force computation.

During the next update of the neighbor list, and as new identities (i.e., indices) are assigned to each atom, the old neighbor-lists are rendered invalid and new ones should be rebuilt. To be more precise, as the **exchange()** routine removes an atom or adds a new one to its $X$ and $V$ arrays, or after the sort routine where the order of atoms changes, the previous indices and consequently the previous neighbor list and border lists become meaningless after these routines are invoked. If the **neighbor_build** and **force_compute** routines run in parallel on two different processors, the processor that runs the **neighbor_build** is actively changing the temporary identities of atoms, while the processor that runs force computation is busy updating the forces for atoms associated with their old identities and effectively invalid neighbor lists. Therefore, the resulting array, $F$, uses obsolete index-atom mappings and produces unusable data for the next iteration. Any argument about algorithm correctness depends on having a way to keep track of atoms index updates on these two processors.

To address this problem, the general structure of our proposed technique is shown in Fig. 10. During the iteration in which the host builds neighbor lists, the BlueField waits until the host finishes the initial integration on the arrays $X$ and $V$. Then, the BlueField begins to read the host's $X$ array and continues with the communication and force computations similar to the non-neighbor-build iterations. But since the order of data has changed in the host, to ensure the correspondence between atoms on the host and the BlueField, we also run the exchange routine on the array $F$, the output of force computation. This step will ensure consistency in the order of atoms between host-exchanged $X$ and $V$ arrays and BlueField-exchanged $F$ array. Additionally, on the host, an array keeps track of orders in the sort routine such that the value stored in index $j$ indicates the new location of the index $j$. The BlueField uses this array in its own sort routine to reorder $F$. Finally, the array $F$ is sent to the host. In preparation for the next neighbor-build iterations, BlueField gets the neighbor lists after the build step completes on the host. In short, achieving more concurrency incurs some overhead, which means the resulting speedups may be less than what one might otherwise predict under ideal overlapped execution.

Overall, our approach is *off-path*, as it uses the BlueField in separated-host mode as an additional compute accelerator, in contrast to an "on-path" packet processing accelerator.

## V. RESULTS

Our experimental evaluation considers the baseline MiniMD implementation and our algorithm (Section IV) when running on the same BlueField-enabled platform from Section III (Table I). This evaluation is organized into three parts. First, we conduct multinode runs where, on each node, we limit concurrency to only one MPI process running on the host and one running on the BlueField, with no other threads of concurrency (i.e., additional MPI or OpenMP processes) within each host or BlueField. This basic assessment aims only to show that co-processor acceleration is feasible. Second, we analyze the accuracy of our method (Section V-2) to confirm that it produces a reasonable numerical result. Third, we revisit the multinode runs and vary the amount of concurrency used on both the host and the BlueField (by varying the number of running OpenMP threads). These results allow us to develop and validate a simple performance model, which helps both to explain our observations and determine other configurations (numbers of host versus BlueField cores) where one might expect performance benefits.

*1) Basic overlap experiments*

In this assessment, the baseline is unmodified MiniMD running on $P \in \{2, 4, 8, 16\}$ nodes with one MPI process per node. This reference is compared against our algorithm, where each host MPI process is paired with one MPI process running concurrently on the BlueField located on that node. For the BlueField-enabled experiments, a host and its paired embedded BlueField process collaboratively complete an assigned job.

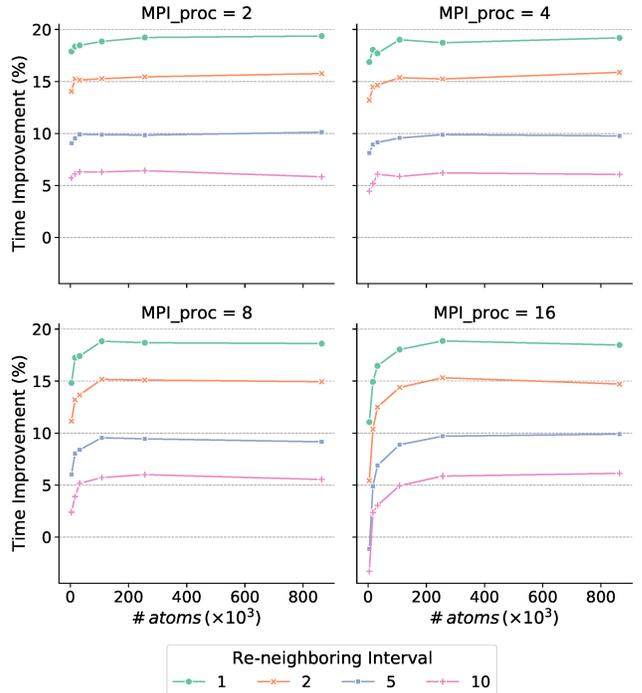

Fig. 11. Up to 20% performance improvements were observed from using our off-path technique compared to the reference algorithm.

We also varied the re-neighboring intervals and numbers of atoms (from 4,000 to 864,000 atoms). The results appear in Fig. 11. The performance improvement is computed as

$$\frac{(\text{total baseline time}) - (\text{total off-path algorithm time})}{\text{total baseline time}}. \quad (2)$$

As shown in Fig. 11, the off-path implementation achieves up to a 20% performance improvement compared to the baseline. This performance boost is unexpected, as the BlueField CPU, acting as a co-processor, is not a high-performance CPU compared to the host CPU cores.

Our experiments also reveal additional trends. For example, as the re-neighboring frequency interval decreases, the percentage of performance improvement increases. This result is expected since the computation overlap between the host and BlueField increases for higher re-neighboring frequencies. Conversely, as the number of atoms per host decreases, the increased overhead of the communication between the host and BlueField outweighs the benefits of overlapped computation between the host and BlueField, leading to limited performance improvements from the off-path algorithm.

We can also estimate the maximum theoretical performance improvement if all communication routines are offloaded to BlueField and these routines on BlueField are completely overlapped with the computation routines on the host. This estimate appears in Fig. 12. The performance improvement values are calculated by $\frac{t_{\text{comm}}}{t_{\text{total}}} \times 100$ using the values from Fig. 7. Comparing the performance gained from our off-path algorithm with the ones shown in Fig. 12, it can be seen that the performance gained by the off-path algorithm is higher for almost all settings, with an absolute difference in performance improvement of up to 16.75%, which indicates the necessity of our algorithmic restructuring.

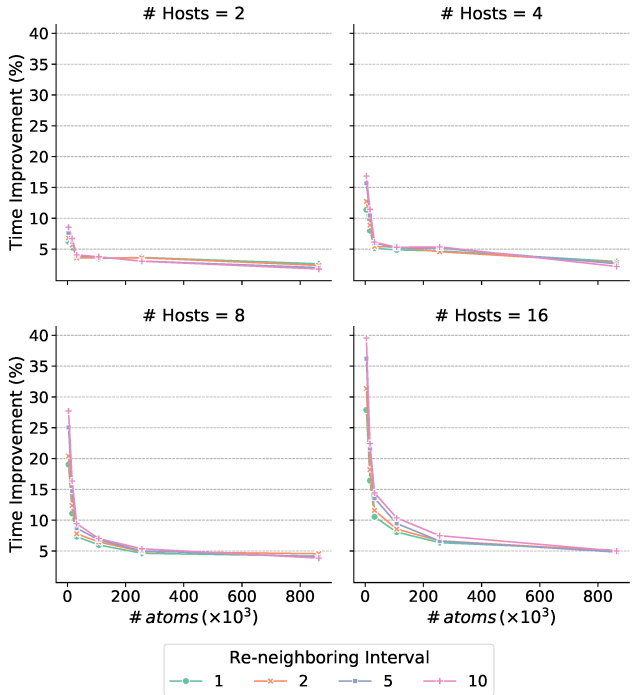

Fig. 12. Theoretical maximum performance improvement possible if all the communication routines are offloaded to BlueField. Relying only on communication offload would tend to underperform compared to the restructured off-path algorithm (cf. Fig. 11).

*2) Accuracy results*

MiniMD reports a variety of statistics about the simulation, including temperature. We use the reported temperature to assess the accuracy of our algorithm through a proxy metric that we call the *temperature divergence rate* (TDR).

To calculate TDR, we start by considering the temperature of the baseline algorithm with highest re-neighboring frequency (i.e., in each iteration) as the *reference temperature*. We calculate the temperature in the intervals of 10 iterations for each experiment and compute the difference between computed temperature in each experiment and the reference temperature in corresponding iterations. Using linear regression, we estimate the rate at which this temperature delta changes as a function of the number of iterations, i.e., $\Delta T(n) = \alpha n + \beta$. The regression coefficient $\alpha$ is the TDR. Ideally, $|\Delta T(n)| \leq \delta$ for some constant $\delta$ if the modified simulation either produced bitwise identical results or incurred errors attributable solely to normal floating-point rounding and nondeterminism during parallel execution. As such, smaller values of $\alpha$ indicate better accuracy.

We compare the TDR by varying re-neighboring intervals for three different skin sizes in Fig. 13. As expected, TDR increases as the re-neighboring interval increases for both baseline and the off-path algorithm. This observation reflects that the probability of moving an atom more than the skin distance increases when the re-neighboring intervals are increased. For the same reason, TDR is higher for smaller skin sizes for the corresponding re-neighboring intervals. Using TDR as a proxy for the accuracy of the algorithm, the results show that the off-path algorithm can be as accurate as the baseline algorithm for a properly selected skin size and re-neighboring interval.

*3) Hybrid MPI/OpenMP performance results*

Our algorithm works best when it can completely hide the force computation time on BlueField by overlapping it with the neighbor build on the host. Imagining future platforms, this degree of achievable overlap will depend on the relative computational power of the host and BlueField device. To study these scenarios, this section considers multinode runs with one MPI process per host and one per BlueField where we also allow the number of cores per host node and cores per BlueField to vary (in contrast to Section V-1). We do so by enabling OpenMP and varying the number of threads. For all the experiments in this section, the re-neighboring interval equals one and the number of atoms is 864,000.

The solid lines in Fig. 14 show the off-path and original MiniMD algorithm runtime for a varying number of OpenMP threads on the host and BlueField. The off-path algorithm's runtime decreases and then stays constant as we increase the number of OpenMP threads on the host. The knee of each curve indicates where the running times of neighbor-build on the host and force-compute on the BlueField are closest. When the number of host threads is less than that at the knee, the neighbor-build on the host is slower than the force-compute on BlueField. Similarly, when the number of host threads is larger than that of the knee, the reverse holds. The knee indicates the maximum number of OpenMP threads on the host where running force computation on BlueField does not slow down the computation.

In contrast to the neighbor build routine, thread synchronization overhead in the force computation routine causes the performance not to scale proportionally to the number of threads. In Fig. 14, the performance with two threads on BlueField drops compared to just one thread, indicating overheads from multithreading. Consequently, the speedup observed from the off-path algorithm in 1 host thread and 1 BlueField thread (compared to 1 host thread with the original algorithm) is not seen in $m$ host threads and $m$ BlueField threads configuration, where $m > 1$.

*4) An explanatory performance model*

To better understand the performance of our offloading algorithm and the results of Fig. 14, we developed a simple performance model that estimates the runtime of the off-path algorithm for different configurations of OpenMP threads on host and BlueField. We run the *original* MiniMD algorithm in both host-only and BlueField-only settings with one MPI process per node and an increasing number of OpenMP threads per node. Based on these experiments, we estimate the required time $t_{\text{force}}$ to complete one **force_compute()** call and the time $t_{\text{neigh}}$ for one **neighbor_build()** call on the host and the BlueField for each configuration. We estimate the theoretical runtime of our algorithm for each configuration as follows, where the superscript notation $p/h/b$ denotes $p$ MPI processes (one per host and one per BlueField) with an additional $h$ OpenMP threads on the host and $b$ OpenMP threads on the BlueField:

$$t_{\text{off-path}}^{p/h/b} = t_{\text{total}}^{p/h/0} - \left\{ t_{\text{force}}^{p/h/0} + t_{\text{neigh}}^{p/h/0} + t_{\text{comm}}^{p/h/0} \right. \\ \left. - \max\left( t_{\text{neigh}}^{p/h/0} + t_{\text{comm}}^{p/h/0}, t_{\text{force}}^{p/0/b} + t_{\text{comm}}^{p/0/b} \right) \right\} \\ \times \frac{\text{\# iterations}}{\text{re-neighboring interval}}. \quad (3)$$

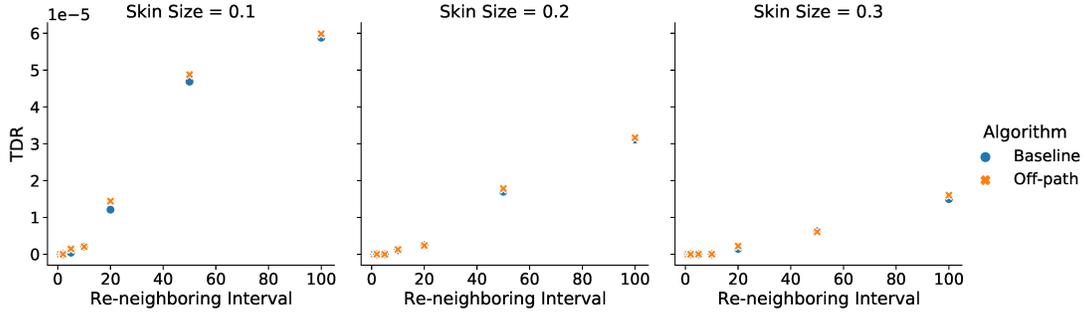

Fig. 13. Off-path algorithm accuracy compared to the reference algorithm. The off-path algorithm gives comparable numerical results.

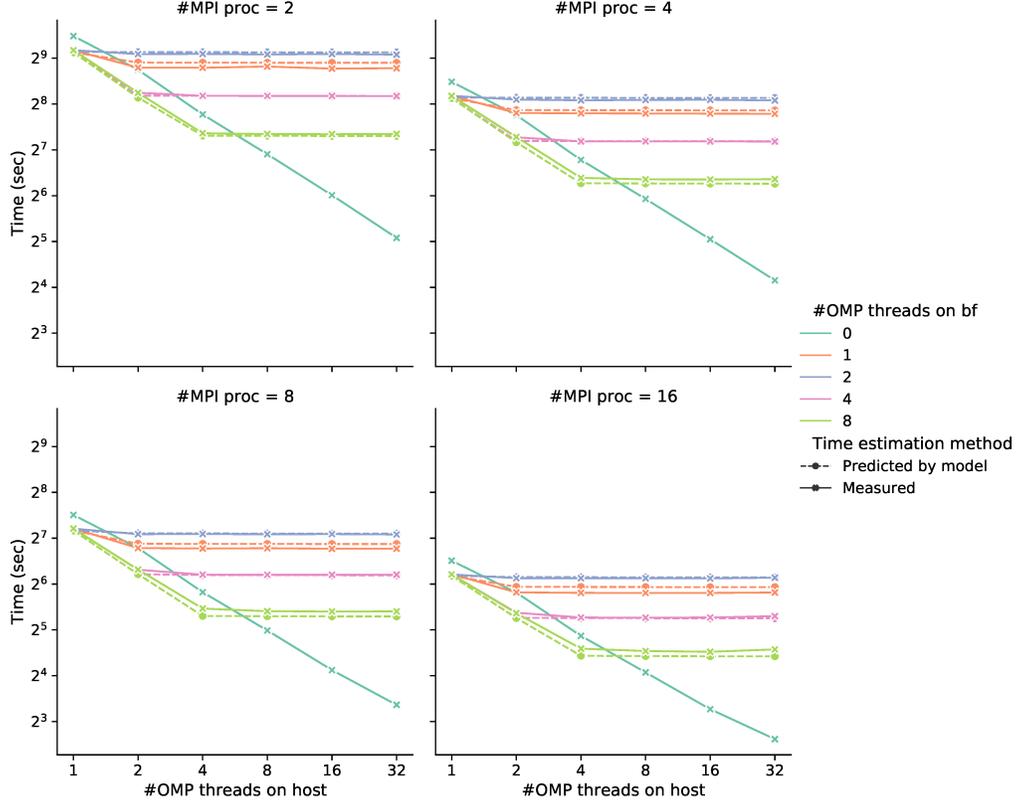

Fig. 14. Performance of the off-path technique for different configurations of host and BlueField. The results indicate regimes where BlueField might benefit performance when fewer host resources (cores) and are consistent with our performance model.

We validate this model in Fig. 14, which compares the estimated time for the off-path algorithm (dashed-line) with the actual measurements (solid line). The model can closely predict the algorithm runtime.

This model also allows us to estimate how the off-path algorithm would perform if we could put more powerful host-like cores on the SmartNIC. In particular, suppose we construct a hypothetical system with the same host CPUs as our system but with a SmartNIC having a 1-core processor comparable to the host core. We refer to this hypothetical SmartNIC as an *auxiliary node*. In Fig. 15, we compare actual measurements for the off-path algorithm in different host/BlueField configurations with the estimated time for the off-path algorithm with this host/auxiliary node. Recall that due to the synchronization overhead in the force computation routine, in the host/BlueField configuration, the performance of the off-path algorithm improves by only 2.5× when increasing the number of OpenMP threads 8-fold. By contrast, our performance model indicates that the off-path algorithm on a system with the hypothetical single-core auxiliary node would perform just as well as the BlueField with eight cores. This match is possible because the synchronization cost does not hinder the performance of the single-thread setup of the auxiliary node.

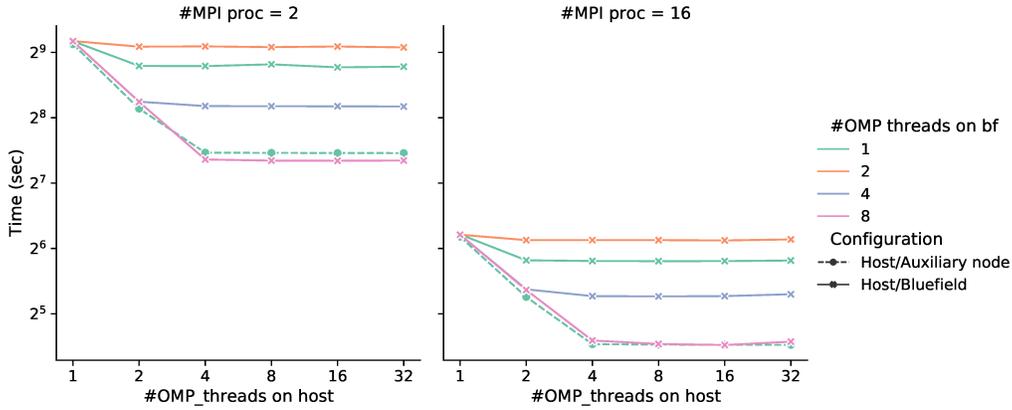

Fig. 15. Performance of the off-path technique on the auxiliary node compared to the BlueField

Based on these observations, we can make several conclusions. The weak cores on the BlueField are more suitable for offloading an application with a high degree of parallelism and low synchronization overhead. If a target application needs more synchronization, it would be fruitful to consider incorporating bigger, fewer cores into the off-path SmartNICs. This approach may also help with communication overhead in these processors. As seen in Section III, the slower CPU on BlueField was the primary reason for the degraded performance in OSU benchmarks, especially for smaller messages (smaller than 16 KiB), for which the CPU is used to copy the buffer data.

## VI. Conclusions and Future Work

Our work shows the viability of off-path computation in an HPC proxy app on a SmartNIC that is accelerated by a low-cost general-purpose multicore processor. This complements the two main bodies of closely related existing work, briefly reviewed in Section II: one considers SmartNICs with specialized ASICs or FPGAs and another considers SmartNICs with general-purpose processors but focuses on accelerating middleware (e.g., MPI communication primitives).

Our study confirms that to achieve performance improvements is likely to require aggressive algorithm and implementation restructuring. This notion had been suggested by others, but not tried, in a BlueField evaluation with the PENNANT proxy app [12]. For MiniMD, we needed to allow for a slight "drift" in terms of neighbor updates. Doing so created a scenario where the BlueField can be used to overlap force computations and neighbor rebuild operations on both the host and SmartNIC. Such algorithmic restructurings hearken back to analogous findings for graphics co-processors. Therefore, although our demonstration is arguably narrow, it indicates other ways in which one might try to use SmartNICs in future studies.

One candidate class of computations for future work are distributed stencils, especially the communication-avoiding variants that use blocking or tiling in time as well as space to trade-off redundant computational work (via larger halo regions) for reduced communication [20]. We suspect these are an algorithmic match to a BlueField-accelerated platform. The redundant work involved might be only a fraction of the main work; therefore, comparatively weak BlueField cores might still be able to execute that redundant work "without cost" to the host. Moreover, since the redundant work is limited to updates in the halo region itself—precisely the data that participates in communication exchanges—there are potential benefits from a lower latency communication path afforded by the SmartNIC. We believe that similar proxy apps, like MiniAMR [21], also have the same characteristics and may benefit from algorithm rewrites to target current and future SmartNIC devices, and we are pursuing these examples.

The commensurate performance improvements we observed for MiniMD appear to be greater than the incremental power increase needed to add a SmartNIC to the host. As a very rough estimate, the Thor servers used for experimental testing consume 316 W on average over a week with a peak power usage of 889 W over the same time period, as measured by standard IPMI logs. A standard InfiniBand adapter might consume 18 W to 20 W on average, and we estimate that the BlueField consumes on average 20 W to 40 W. In context, an increase in average system power ([20–40]/316) of 6% to 13% enabled our speedups of up to 20%, an encouraging result.

Nevertheless, the BlueField platform itself is also arguably limited (Section III), which opens numerous redesign opportunities. One is the hypothetical "auxiliary node" discussed in Section V-4, which has just a single, albeit more powerful, host-like core while matching what the 8-core Arm achieves. But other "rebalancings" of processor configurations may be possible, too. For instance, some of the computation that we had assumed would be most profitable to offload onto the BlueField in our new off-path algorithm, like sorting and basic analysis or near-communication data structure reorganization, was severely limited by the low core-performance of the embedded Arm cores. But the aggregate efficiency of BlueFields in standalone mode was also strong when matching the host on peak performance (Section III-D). So, another possible architecture could rebalance the ratio of lightweight BlueField processing elements and more heavyweight cores, with BlueField nodes running easily parallelized and NEON-vectorized operations and "offloading" irregular computations onto the heavyweight cores, similar to early suggestions about GPGPUs.

Thinking more broadly, understanding these opportunities and limitations fully will require additional work, to include more case studies, performance modeling, and programming model improvements. Performance modeling would help identify when offload is profitable or what hardware parameters would need improvement to deliver a benefit. Programming model are needed to improve productivity; in our case, we had to "hack" the code in an ugly way to construct the separate control paths for force computation and neighbor list updates. New programming models that simplify how offload is implemented would facilitate future experiments.

ACKNOWLEDGEMENTS

This paper describes objective technical results and analysis. Any subjective views or opinions that might be expressed in the paper do not necessarily represent the views of the U.S. Department of Energy or the United States Government. This work has been funded through Sandia National Laboratories (Contract Number 2200840). Sandia National Laboratories is a multi-mission laboratory managed and operated by National Technology and Engineering Solutions of Sandia, LLC., a wholly owned subsidiary of Honeywell International, Inc., for the U.S. Department of Energy's National Nuclear Security Administration under contract DE-NA0003525. We thank the HPC-AI Advisory Council testbed for BlueField testing and evaluation. This work also used resources from the CRNCH Rogues Gallery testbed (NSF CNS #2016701).